\documentstyle[pra,aps]{revtex}
\begin{document}
\renewcommand{\theequation}{\thesection-\arabic{equation}}

\title{SOLUTION OF SRS ON THE FINITE INTERVAL
\footnote{ Submitted to Physics Letters A \hfill Preprint \#PM 95-38}}
\author{ J. LEON \\
Physique Math\'ematique et Th\'eorique, CNRS-URA 768\\
34095 MONTPELLIER Cedex 05 (FRANCE)}\maketitle


\begin{abstract}
The equations of transient stimulated Raman scattering on the 
finite interval are solved by the spectral transform method on the
{\em semi-line}.  As the problem has a free end, the pump and Stokes output at 
{\em finite distance} can be constructed as the solution of a linear Cauchy-Green
integral equation. 
\end{abstract}

\section{Introduction}
Since the first observation in 1983 of the creation of a {\em Raman spike} in the
pump depletion zone by  Dr\"uhl, Wenzel and Carlsten \cite{druwen}, there has been
a large amount of studies of stimulated Raman scattering (SRS) of long laser
pulses in gas.
The reason is that the Raman spike has been shown to be the macroscopic
manifestation of large fluctuations  of the phase of the initial Stokes
wave. Raman spike generation were then predicted \cite{engbow},
have been given a coherent-mode description \cite{rayliwam}, and
experiments have been preformed where SRS grows spontaneously on initial
fluctuations \cite{macswa2}. The Raman spike hence appears as a means to study
the  quantum  properties of the  Stokes wave initiation, which gives
informations on the {\em phase of the electromagnetic vacuum}.
This comes in addition to the process of Stokes growth which amplifies the
quantum fluctuations of the medium. A general discussion of the {\em quantum
coherence properties of SRS} is given in \cite{raywal}.

As the SRS equations possess a Lax pair \cite{chusco}, there has been
many attemps to modelize the Raman spike as a soliton (see e.g.
\cite{soliton1}\cite{soliton2}). But it has been proved recently that actually 
the Raman spike occurs in the spectral transform theory as a manifestation 
of the continuous spectrum  (hence it is not a soliton) when
for a short period of time the {\em reflection coefficient}  becomes close to zero
\cite{leon-prl}. These results were obtained by solving the initial-boundary value
problem for the SRS equations on the infinite line, which, from a physical point
of view, is inconsistent with a finite dephasing time of the medium oscillators.
However the results obtained are strikingly close to the experimental data
\cite{leon-pra}. Here we construct a solution of SRS on the finite interval by
using the spectral transform on the semi-line, first proposed in the context of
{\em nonlinear polarization dynamics} \cite{leon-sasha}.

The interaction of light with a material medium, in the case when
a laser pump pulse (frequency $\omega_L$, envelope $A_L$)
interacts with the optical phonons (eigenfrequency $\omega_V$, envelope $Q$)
to give rise to a down-shifted laser  pulse 
(Stokes emission, frequency $\omega_S$, envelope $A_S$) according to the selection
rules ($\Delta\omega$ is the  the detuning 
from the Raman resonance, $\Delta k$ is the phase missmatch)
\begin{equation}
\omega_S=\omega_L-\omega_V-\Delta\omega,\quad
k_S=k_L-k_V-\Delta k
\label{selec-rules}\end{equation}
can be modelized by the following {\em slowly varying envelope approximation}
\cite{yariv}
\begin{eqnarray}
({\partial\over\partial Z}+{\eta\over c}{\partial\over\partial T}) A_L
&=&\frac{i}{4}{N\alpha'_0\over\eta c}\sqrt{\omega_S\omega_L}\ Q A_S\
\exp[i(\Delta k Z-\Delta\omega T)],\nonumber\\
({\partial\over\partial Z}+{\eta\over c}{\partial\over\partial T}) A_S
&=&\frac{i}{4}{N\alpha'_0\over\eta c}\sqrt{\omega_S\omega_L}\ Q^*\  A_L\
\exp[-i(\Delta kZ-\Delta\omega T)],
\label{base-physique}\\
{\partial\over\partial T}Q+\frac{1}{T_2}Q
&=&\frac{i}{4}{\epsilon_0\alpha'_0\over
m\omega_V}\sqrt{\frac{\omega_S}{\omega_L}}\ A_L   A_S^*\
\exp[-i(\Delta kZ-\Delta\omega T)].
\nonumber\end{eqnarray}
The electromagnetic wave field $E(Z,T)$ 
and the material excitation $\tilde X(Z,T)$ are obtained as
\begin{equation}
E(Z,T)=\frac12
A_L\ \exp[i(k_LZ-\omega_LT)]+\frac12\sqrt{\frac{\omega_S}{\omega_L}}
A_S\ \exp[i(k_SZ-\omega_ST)]+c.c.
\label{champ-E}\end{equation}
\begin{equation}
\tilde X(Z,T)=\frac12 Q\ \exp[i(k_VZ-\omega_VT)]+c.c.
\label{champ-X}\end{equation}
Hereabove $\alpha'_0$ is the differential
polarizability at equilibrium, $c/\eta$ is the light velocity in the medium,
$N$ is the density of oscillators of mass $m$,
$T_2$ is the relaxation time of the medium.

For a medium initially in the ground state
\begin{equation}
Q(Z,0)=0,
\label{init-medium}\end{equation}
and for an arbitrary set of input pump and Stokes envelopes profiles
(for any value of the missmatch $\Delta k$)
\begin{equation}
A_L(0,T)=I_L(T-\frac\eta cZ),\quad A_S(0,T)=I_S(T-\frac\eta cZ),
\label{bound-laser}\end{equation}
we give here the  output values of both light waves profiles
$A_L(L,T)$ and $A_S(L,T)$ in terms of the solution of a Cauchy-Green linear
integral equation.

The solution, constructed by the inverse spectral transform 
theory (IST), is actually exact for {\em infinite relaxation times},
and we have proposed an approximate
solution for finite $T_2$ which matches the experiments with high accuracy
\cite{leon-pra}.
The method applies also for a non-vanishing initial state of the medium but the
solution in this case is in general not explicit (this is relevant in
physical situations where the quantum fluctuations of the initial population
density of the two-level medium is taken into account).

The fact that IST can be applied to SRS on the finite interval has been first
proposed by Kaup in 1983 \cite{soliton2}. However the evolution of the spectral
data given there does not correspond to the boundary problem (\ref{bound-laser})
and in particular, as this evolution is {\em homogeneous}, it does not allow for
the growth of the Stokes seed on a medium initially at rest. In a different
context, a {\em nonhomogeneous} evolution of the spectral data has been obtained
in \cite{gabzak} where the self-induced transparency equations are solved for an
arbitrary {\em initial} ($t=-\infty$) population density of the two-level
medium. IST has been later used to solve an initial-boundary value problem on
the half-line for the nonlinear Schr\"odinger equation (NLS) by Fokas and Its
\cite{fokas},  but in this case the required boundary data in $x=0$ 
for the potential itself renders {\em nonlinear} the  evolution of the spectral 
transform.

The property of SRS of being solvable on the finite interval results simply from
the nature of the equations  for which the
initial-boundary value problem (\ref{init-medium})(\ref{bound-laser}) is well
posed and does not require new constraints when it is given on the finite
interval (this is not so for NLS for which the vanishing boundary values 
at infinity become some prescribed boundary value in $x=0$). Consequently the
method applies for every other case of solvable evolutions with nonanalytic
dispersion relations when precisely passing to the finite interval does not
imply adding information or constraint. However we will discover that there are
requirements of {\em analyticity of the boundary data} in order for the problem
to be integrable.

Before going to the method of solution, it is convenient to rescale the system
(\ref{base-physique}) into a {\em dimensionless} system, which
goes with defining the new  variables
\begin{equation}
x= \frac1L Z,\quad t=\frac cL(T-\frac\eta c Z).
\label{coordi}\end{equation}
Then the dimensionless rescaled material excitation is defined as
(the differential polarizability $\alpha'_0$ has the dimension of a surface)
\begin{equation}
q(x,t)=\frac i4\ \frac{N\alpha_0'}{\eta c}\sqrt{\omega_L\omega_S}\ LQ(Z,T),
\label{rescale-Q}\end{equation}
while $A_L$ ans $A_S$ are rescaled by using the boundary conditions as
\begin{equation}
a_L(k,x,t)=\frac{A_L(\Delta k,Z,T)}{I_m},\quad 
a_S(k,x,t)=\frac{A_S(\Delta k,Z,T)}{I_m},\quad 
I_m=\displaystyle{\mathop{\mbox{max}}_{T>0}} |I_L|,
\label{rescale-E}\end{equation}
which scales to 1 the incident laser pump amplitude. 
The dependence on the phase missmatch is represented here by the
dimensionless wave number
\begin{equation}
k=\frac12 L(\Delta k-\frac\eta c\Delta\omega),
\label{essential}\end{equation}
which we shall refer to as the {\em essential missmatch parameter}.
Finally, for
\begin{equation}
g_0=\frac{L^2}{16}{N\alpha'_0\epsilon_0\over\eta mc^2}
{\omega_S\over\omega_V}\ I_m^2,
\label{qdyn}\end{equation}
(dimensionless), 
the system (\ref{base-physique}) becomes for infinite dephasing time
\begin{equation}
\partial_x a_L=q  a_S\ e^{-i\Delta\omega t}\ e^{2ikx},
\quad
\partial_x a_S=-q^* a_L\ e^{i\Delta\omega t}\ e^{-2ikx},
\label{maxsimp}\end{equation}
\begin{equation}
q_t= - g_0  a_L a_S^*\ e^{i\Delta\omega t}\
e^{-2ikx}.\label{qinteq-nu}\end{equation}

Due to the dependence of the field envelopes $a_L$ and $a_S$ on the essential
missmatch $k$, it is necessary to consider the cooperative interaction of all
$k$-components with the medium.
Then we need actually to consider the medium excitation as resulting from
all $k$ components and replace (\ref{qinteq-nu}) with
\begin{equation}
q_t= - g_0\int dk  a_L a_S^*\ e^{i\Delta\omega t}\
e^{-2ikx},
\label{qinteq}\end{equation}
where now the input ($x=0$) values of the pump wave $a_L$ and the Stokes 
wave $a_S$ are also function of $k$ sharply distributed
around $k=0$ (the resonnance), that we denote by
\begin{equation}
a_L(k,0,t)=J_L(k,t),\quad a_S(k,0,t)=J_S(k,t).
\label{in-A}\end{equation}

We will demonstrate that the system (\ref{maxsimp})(\ref{qinteq}), with the  
initial data  $q(x,0)$ in $L^1$ and boundary values (\ref{in-A}), where
$J_L(k,t)$  (resp. $J_S(k,t)$) has an analytic continuation in ${\rm Im}(k)>0$
(resp. ${\rm Im}(k)<0$) vanishing as $|k|\to\infty$, is integrable on the
semi-line $x>0$. Moreover the solution furnishes the output values of the pump
and Stokes waves in $Z=L$ (i.e. $x=1$), which gives the solution of the SRS
equations on the finite interval.

\setcounter{equation}{0}
\section{The spectral problem on the semi-line}

\paragraph*{Basic definitions.}
We briefly give the basic notions on the
Zakharov-Shabat spectral problem on the semi-line
for the $2\times2$ matrix $ \nu(k,x,t)$ in the potentials $q(x,t)$ and $r(x,t)$
\begin{equation}
\nu_x+ik[\sigma_3,\nu]=\left(\matrix{0&q \cr r&0}\right)\nu,
\quad x\ge0,\quad  t\ge0.
 \label{zs}\end{equation}
The solution $\nu$ can be verified to obey
\begin{equation}
{\partial\over\partial x}\det\{\nu\}=0.
\label{det-nu}\end{equation}
Two fundamental solutions, say $\nu^\pm$, are
defined by 
\begin{equation}\left(\matrix{\nu_{11}^+ \cr\nu_{21}^+ \cr}\right)=
\left(\matrix{1\cr0\cr}\right)+
\left(\matrix{\int_{0}^{x}dx' q \nu_{21}^+ \cr
      \int_{0}^{x}dx' r \nu_{11}^+ e^{2ik(x-x')}\cr}
       \right)                          \label{int-1}\end{equation}
\begin{equation}\left(\matrix{\nu_{12}^+ \cr\nu_{22}^+ \cr}\right)=
\left(\matrix{0\cr1\cr}\right)+
\left(\matrix{-\int_{x}^{\infty}dx' q \nu_{22}^+ 
                              e^{-2ik(x-x')}\cr
     \int_{0}^{x}dx' r \nu_{12}^+ \cr}
       \right)                        \label{int-2}\end{equation}
 \begin{equation}\left(\matrix{\nu_{11}^- \cr\nu_{21}^- \cr}\right)=
\left(\matrix{1\cr0\cr}\right)+
\left(\matrix{\int_{0}^{x}dx' q \nu_{21}^- \cr
     -\int_{x}^{\infty}dx' r \nu_{11}^- e^{2ik(x-x')}\cr}
       \right)                               \label{int-3}\end{equation}
\begin{equation}\left(\matrix{\nu_{12}^- \cr\nu_{22}^- \cr}\right)=
\left(\matrix{0\cr1\cr}\right)+
\left(\matrix{\int_{0}^{x}dx' q \nu_{22}^- 
                              e^{-2ik(x-x')}\cr
     \int_{0}^{x}dx' r \nu_{12}^- \cr} \right)
\label{int-4}\end{equation}

\paragraph*{Riemann-Hilbert problem.}
In order to derive a well-posed Riemann-Hilbert
problem for $\nu(k)$, we require that $\nu(k)$ be bounded for
large $k$ and hence define it as
\begin{equation}
\nu(k,x,t)=\left\{\matrix{\nu^+(k,x,t), \quad {\mbox{Im}}(k)>0 \cr
           \nu^-(k,x,t),\quad {\mbox{Im}}(k)<0}\right\}.
\label{nu-def}\end{equation}
Indeed it obeys 
\begin{equation}
k\to\infty\quad:\quad\nu(k)\sim 1+{\cal O}({1\over k}),
\label{k-inf}\end{equation}
which is obtained from the integral equations by integration by part. 
The scattering coefficients (functions of $k$ and parametrized by $t$) 
are defined for $k$ real, as 
\begin{equation}
\rho^+=-\int_0^\infty dx'\ q\nu_{22}^+ e^{2ikx'},\quad
\rho^-=-\int_0^\infty dx'\ r\nu_{11}^- e^{-2ikx'},
\label{rho+-}\end{equation}
\begin{equation}
\tau^+=1+\int_0^\infty dx'\ r\nu_{12}^+,\quad
\tau^-=1+\int_0^\infty dx'\ q\nu_{21}^-.
\label{tau+-}\end{equation}
Following standard methods (see e.g. \cite{leon-jmp}, especially the appendix),
one can prove from the above integral equations the the solution $\nu$ obeys on
the real axis, the following Riemann-Hilbert problem
\begin{equation}
\nu_1^+ - \nu_1^- = -e^{2ikx}\rho^-\nu_2^-, \quad
\nu_2^+ - \nu_2^- =  e^{-2ikx}\rho^+\nu_1^+. 
\label{R-H}\end{equation}
The method consists simply in comparing e.g. the integral equations for 
$(\nu_1^+ - \nu_1^-)e^{-2ikx}$ to the one for $\rho^-\nu_2^-$.

\paragraph*{Bound states.}
As the integrals run on the finite support $[0,x]$, 
the solutions $\nu^+_1$ and $\nu^-_2$ 
of the Volterra integral equations (\ref{int-1})(\ref{int-4}) are analytic
(entire functions of $k$)
in the complex plane. The property (\ref{det-nu}) is used to compute the
determinants of $(\nu^+_1,\nu^+_2)$ and of $(\nu^-_1,\nu^-_2)$ both at $x=0$ and
$x=\infty$, which gives
\begin{equation}
\nu^+_{11}(k,\infty,t)={1\over\tau^+(k,t)},\quad
\nu^-_{22}(k,\infty,t)={1\over\tau^-(k,t)}.
\label{inv-tau}\end{equation}
Hence the quantity $1/\tau^+$ ($1/\tau^-$) is an entire functions of $k$
and can have a number $N^+$
($N^-$) of zeroes $k_n^+$ ($k_n^-$) which, for bounded $r$ and $q$ are simple an
of finite number. 

Consequently, the solutions  $\nu^-_1$ and $\nu^+_2$  of the Fredholm integral
equations (\ref{int-3}) and (\ref{int-2}), which can be written
also 
\begin{equation}\left(\matrix{\nu_{11}^- \cr\nu_{21}^- \cr}\right)=
\left(\matrix{1\cr0\cr}\right)\tau^- -
\left(\matrix{\int_{x}^{\infty}dx' q \nu_{21}^- \cr
     \int_{x}^{\infty}dx' r \nu_{11}^- e^{2ik(x-x')}\cr}\right) ,  
\label{int-3-bis}\end{equation}
\begin{equation}\left(\matrix{\nu_{12}^+ \cr\nu_{22}^+ \cr}\right)=
\left(\matrix{0\cr1\cr}\right)\tau^+ -
\left(\matrix{\int_{x}^{\infty}dx' q \nu_{22}^+e^{-2ik(x-x')}\cr
                    \int_{x}^{\infty}dx' r \nu_{12}^+ \cr}\right),
\label{int-2-bis}\end{equation}
have respectively the $N^-$ and $N^+$ simple poles $k_n^-$ and $k_n^+$
of $\tau^-$ and $\tau^+$. Then the integral equations for the residues allow to
obtain
\begin{equation}
\displaystyle\mathop{\rm Res}_{k_n^-}\ \nu_1^-=
C_n^-\nu_2^-(k_n^-)\exp[2ik_n^-x],\quad
C_n^-=\displaystyle\mathop{\rm Res}_{k_n^-}\ \rho^-,
\label{apres2}\end{equation}
\begin{equation}
\displaystyle\mathop{\rm Res}_{k_n^+}\ \nu_2^+=
C_n^+\nu_1^+(k_n^+)\exp[-2ik_n^+x],\quad
C_n^+=\displaystyle\mathop{\rm Res}_{k_n^+}\ \rho^+.
\label{apres1}\end{equation}
As for the Riemann-Hilbert relations (\ref{R-H}), the method consists in comparing
the integral equations for $\exp[-2ik_n^-x]{\rm Res}\ \nu_1^-$ and for
$C_n^-\nu_2^-(k_n^-)$.

\paragraph*{Boundary behaviors.}
An important relation for the following is the so-called {\em unitarity} 
relation
\begin{equation}
\tau^+\tau^-=1-\rho^+\rho^-,
\label{unitarity}\end{equation}
obtained by computing $\nu^-_{22}(k,\infty,t)$ by using (\ref{R-H}) and by
comparing the result with (\ref{inv-tau}).

Consequently the boundary behaviors of $\nu$ can be written 
\begin{equation} x=0\quad :\quad
\nu^+=\left(\matrix{1 & \rho^+ \cr 0 & 1}\right),\quad
\nu^-=\left(\matrix{1 & 0 \cr \rho^-  & 1}\right),
\label{bound-x=0}\end{equation}
\begin{equation}x=\infty\quad :\quad
\nu^+=\left(\matrix{1/\tau^+ & 0\cr -e^{2ikx}\rho^-/\tau^- & \tau^+}\right),
\quad
\nu^-=\left(\matrix{\tau^- & -e^{-2ikx}\rho^+/\tau^+ \cr 0  & 1/\tau^-}\right).
\label{bound-x=a}\end{equation}

\paragraph*{The $\bar\partial$ problem.} 
The analytic properties of $\nu$ can be summarized in the following
formula 
\begin{equation}
{\frac{\partial\nu}{\partial\bar k}}=\nu\ R \exp[2ik\sigma_3x],
\label{d-bar}\end{equation}
with the {\em spectral transform} 
\begin{equation}
R={i\over2}
\left(\matrix{0 & \rho^+\delta^+ \cr
             -\rho^-\delta^-& 0} \right)-
2i\pi\left(\matrix{0&\sum C_n^+ \delta(k-k_n^+)\cr
            \sum C_n^- \delta(k-k_n^-) & 0} \right),
 \label{R-str}\end{equation}
where the distributions $\delta^\pm$   and
$\delta(k-k_n)$ are defined as 
\begin{eqnarray} 
\int\!\!\!\int d\lambda \wedge d\bar{\lambda} f(\lambda)
\delta^{\pm}(\lambda_I)  &&= - 2i \int^{+\infty}_{-\infty} d\lambda_R
f(\lambda_R \pm i0),\nonumber\\
\int\!\!\!\int d\lambda \wedge d\bar{\lambda} f(\lambda)\delta(\lambda-k_n)&&=
f(k_n)
\label{deltapm} \end{eqnarray}
with the notation $\lambda=\lambda_R+i\lambda_I$.

\setcounter{equation}{0}
\section{Inverse spectral problem}

\paragraph*{The Cauchy-Green integral equation.}
The inverse problem is solved by integrating the
$\bar\partial$-equation (\ref{d-bar}) with the boundary (\ref{k-inf}).
We prove here the following theorem: 
the solution $f(k,x,t)$ of the Cauchy-Green integral equation
\begin{equation}
{f}(k,x,t)={\bf 1}+{1\over2i\pi}\int\!\!\!\int
\frac{d\lambda\wedge d\bar\lambda}{\lambda-k}
\;\;{f}(\lambda,x,t)R(\lambda,t)\exp[2i\lambda\sigma_3x].
\label{cauchy}\end{equation}
coincides with the solution $\nu(k,x,t)$ of (\ref{int-1})-(\ref{int-4}) if
the reflection coefficient possess a meromorphic continuation
with simple poles $k_n^\pm$ and residues $C_n^\pm$, namely
\begin{equation}\label{rho-analytic}
\pm{\rm Im}(k)>0\ \Rightarrow\ {\partial\over\partial\bar k}\ \rho^\pm(k)=
2i\pi\sum C_n^\pm\ \delta(k-k_n^\pm).
\end{equation}
Such analytical property of the spectral data occur when the potentials are on
compact support. Therefore we shall assume here that the {\em initial datum}
$q(x,0)$ is indeed on compact support  and we will have to demonstrate that the
time evolution conserves this analytical property.
Actually the physically interesting case is when $q(x,0)$  vanish identically.
This implies no bound states which will be assumed in the following
(everything can be extended to the case with bound states
as they appear as the poles $k_n^+$ of $\rho^+$ in 
${\rm Im}(k)>0$, and those ($k_n^-$) of $\rho^-$ in ${\rm Im}(k)<0$).

\paragraph*{Proof of the Theorem.}
The first step is to verify from (\ref{d-bar}) that the function
${f}$ is solution of the differential problem (\ref{zs}). This is easily done by 
following the method  of \cite{leon-jmp}, in short: prove the relation
\begin{equation}
{\frac{\partial}{\partial\bar k}}
[({f}_x-ik{f}\sigma_3){f}^{-1}]=0
\label{bas-dbar-x}\end{equation} 
and integrate it. Then the solution ${f}$ of (\ref{cauchy}) solves
the spectral problem (\ref{zs}) with
\begin{equation}
\left(\matrix{0&q \cr r&0\cr}\right)=i[\sigma_3,{f}^{(1)}]
\label{pot-nu}\end{equation}
where ${f}^{(1)}$ is the coefficient of $1/k$ in the Laurent expansion of 
${f}(k)$.

The second step consists in proving
\begin{equation}
{f}_1^+(k,x,t)=\nu_1^+(k,x,t),\quad {f}_2^-(k,x,t)=\nu_2^-(k,x,t),
\label{first-analytic}\end{equation}
which is acheived just by comparing the values of these vectors in $x=0$.
The functions ${f}^+$ and ${f}^-$ solve the following
coulped vectorial system (forget for a while the $(x,t)$-dependence) on the
real axis ${\mbox{Im}}(k)=0$:
\begin{equation}
{f}_1^+(k)=\left(\matrix{1\cr0}\right)-{1\over2i\pi}\int
{d\lambda\over\lambda-(k+i0)}
\  \rho^-(\lambda){f}_2^-(\lambda)e^{2i\lambda x},
\label{cauchy-1+}\end{equation}
\begin{equation}
{f}_2^-(k)=\left(\matrix{0\cr1}\right)+{1\over2i\pi}\int
{d\lambda\over\lambda-(k-i0)}
\  \rho^+(\lambda){f}_1^+(\lambda)e^{-2i\lambda x}.
\label{cauchy-2-}\end{equation}
Now, by closing the contour of
integration in ${\rm Im}(\lambda)<0$ for ${f}_1^+$ and 
${\rm Im}(\lambda)>0$ for ${f}_2^-$, the Cauchy theorem for the
analyticity requirement (\ref{rho-analytic}) leads to
\begin{equation}
\forall x<0\ ,\ {\rm Im}(k)=0\ :\
{f}_1^+(k,x,t)=\left(\matrix{1\cr0}\right),\quad
{f}_2^-(k,x,t)=\left(\matrix{0\cr1}\right).
\label{val-0-1}\end{equation}
Consequently, the two matrices $({f}_1^+,{f}_2^-)$ and $(\nu_1^+,\nu_2^-)$
solve the same first order differential problem and have the same values in
$x=0$, so (\ref{first-analytic}) is proved.

The third step results in proving
\begin{equation}\label{second-analytic}
{f}_1^-(k,x,t)=\nu_1^-(k,x,t),\quad
{f}_2^+(k,x,t)=\nu_2^+(k,x,t),
\end{equation}
which is performed by evaluating the functions
\begin{eqnarray}\label{f-infty}
&&{f}_{12}^+={1\over2i\pi}\int{d\lambda\over\lambda-(k+i0)}\
\rho^+{f}_{11}^+e^{-2i\lambda x}\nonumber\\
&&{f}_{21}^-=-{1\over2i\pi}\int{d\lambda\over\lambda-(k-i0)}\
\rho^-{f}_{22}^-e^{2i\lambda x}
\end{eqnarray}
as $x\to\infty$. A usefull formula is
\begin{equation}\label{lim-distr}
\lim_{x\rightarrow \pm\infty }P\!\!\!
\int\frac{d\lambda}{\lambda-k}\ e^{i(\lambda-k)x}f(\lambda)
=\pm i\pi f(k),
\end{equation}
where $P\!\!\int$ denotes the Cauchy principal value integral, and the
Sokhotski-Plemelj formula
\begin{equation}\label{sokhotski}
\int\frac{d\lambda}{\lambda-(k\pm i0)}\ f(\lambda)
=\pm i\pi f(k)+P\!\!\!\int\frac{d\lambda}{\lambda-k}\ f(\lambda).
\end{equation}
Since it has been already proved that ${f}_1^+=\nu_1^+$ and
${f}_2^-=\nu_2^-$, the functions ${f}_{11}^+$ and ${f}_{22}^-$ have from
(\ref{int-2}) and (\ref{int-3}) bounded behaviors as $x\to\infty$. Consequently
\begin{equation}\label{val-infini-1}
\lim_{x\to+\infty}{f}_{12}^+=0,\quad
\lim_{x\to+\infty}{f}_{21}^-=0.
\end{equation}
The remaining functions ${f}_{11}^-$ and ${f}_{22}^+$ are  obtained 
by using the differential form of (\ref{cauchy}), namaly
\begin{equation}
{f}_{11}^+ - {f}_{11}^- = -e^{2ikx}\rho^-{f}_{12}^-, \quad
{f}_{22}^+ - {f}_{22}^- =  e^{-2ikx}\rho^+{f}_{21}^+. 
\label{R-H-f}\end{equation}
With (\ref{val-0-1}), this gives
\begin{equation}\label{val-0-2}
\forall x<0\ ,\ {\rm Im}(k)=0\ :\ {f}_{11}^-(k,x,t)=1,\quad 
{f}_{22}^+(k,x,t)=1.
\end{equation}
Finally, the two matrices $({f}_1^-,{f}_2^+)$ and $(\nu_1^-,\nu_2^+)$
solve the same first order differential problem and have the same values in
$x=0$ for the diagonal elements and in $x=\infty$ for the off-diagonal, hence
they are equal.
The theorem is proved and an immediate consequence is that
the coefficients $\rho^\pm$ are effectively given by (\ref{rho+-}).

From now on we will use the single matrix $\nu$ as denoting the common
solution to (\ref{cauchy}) and (\ref{int-1})-(\ref{int-4}).

\paragraph*{Reduction.}
It is convenient for physical applications to
assume the following {\em reduction}
\begin{equation}
r=- \bar q
\label{reduc}\end{equation}
for which 
\begin{equation}
(\nu_{11}^\pm)^*=\nu_{22}^\mp,\quad (\nu_{12}^\pm)^*=-\nu_{21}^\mp,
\label{red-nu}\end{equation}
\begin{equation}
\rho^+=-(\rho^-)^*,\quad \tau^+=(\tau^-)^*,\quad
k_n^+=\overline{k}_n^-,\quad C_n^+=\overline{C}_n^-. 
 \label{red-struc}\end{equation}
The  overbar denotes the complex conjugation and the ``star''
\begin{equation}\label{star}
f^*(k)=\overline{f\left(\bar k\right)}.\end{equation}

As a consequence the boundary values of $\nu$ become 
\begin{equation}x=0\ :\ 
\nu^+=\left(\matrix{1 & \rho \cr 0 & 1}\right),\quad
\nu^-=\left(\matrix{1 & 0 \cr -\rho^*  & 1}\right),
\label{bound-0}\end{equation}
\begin{equation}x=\infty\ :\ 
\nu^+=\left(\matrix{1/\tau & 0\cr e^{2ikx}\rho^*/\tau^* & \tau}\right),\quad
\nu^-=\left(\matrix{\tau^* & -e^{-2ikx}\rho/\tau \cr 0  & 1/\tau^*}\right),
\label{bound-a}\end{equation}
where we use the notation $\rho=\rho^+$ and $\tau=\tau^+$, for real $k$.

\paragraph*{Transmission coefficient.} 
It is usefull for the following to derive
the relationship between $\tau$ and $\rho$. Although this can be done in
general, we still consider only the case when no bound states are present (this
will be the case of interest). From the definitions (\ref{tau+-}), we have for
large  $k$
\begin{equation}
\tau^\pm(k)\sim 1+{\cal O}({1\over k}).
\label{behav-tau}\end{equation}
Defining then ${h}(k)$ as $\ln(\tau^+)$ for ${\mbox{Im}}(k)\ge0$, and
$-\ln(\tau^-)$ for ${\mbox{Im}}(k)\le0$,
its discontinuity on the real $k$-axis can be written by means of 
(\ref{unitarity})
\begin{equation}
{h}^+-{h}^-=\ln(1-\rho^+\rho^-).
\label{disc-f}\end{equation}

As  from (\ref{behav-tau}), ${h}(k)$ vanishes for large $k$, the above
Riemann-Hilbert problem has the following solution 
\begin{equation}
{\rm Im}(k)\ne0\ :\ {h}(k)={1\over2i\pi}\int{d\lambda\over\lambda-k}\ 
\ln(1-\rho^+\rho^-),
\label{sol-f}\end{equation}
and hence $\tau^\pm$ are given from $\rho^\pm$. In the reduction (\ref{reduc}),
this relation can be written
\begin{equation}
\tau=\sqrt{1+|\rho|^2}\ e^{i\theta},\quad
\theta=-{1\over2\pi}P\!\!\!\!\int{d\lambda\over\lambda-k}\ \ln(1+|\rho|^2).
\label{tau-rho}\end{equation}

\setcounter{equation}{0}
\section{Lax pair and general evolution}

\paragraph*{Lax pair.}
The compatibility 
\begin{equation}
U_t-V_x+[U,V]=0
\label{UV}\end{equation}
between the two following spectral problems
\begin{equation}
\nu_x=\nu\ ik\sigma_3+U\nu,\quad U=-ik\sigma_3+\left(\matrix{0&q \cr
r&0}\right),
\label{lax1}\end{equation}
\begin{equation}
\nu_t=\nu\Omega+V\nu,\quad V={1\over2i\pi}\int\!\!\!\int
\frac{d\lambda\wedge d\bar\lambda}{\lambda-k}\ \nu(Me^{2i\lambda\sigma_3x}
-{\partial\Omega\over\partial
\bar\lambda})\nu^{-1},
\label{lax2}\end{equation}
leads to the evolutions
\begin{equation}
{\partial\over\partial t}\left(\matrix{0&q \cr r&0}\right)=
-{1\over2\pi}\int\!\!\!\int[\sigma_3\ ,\ \nu(Me^{2i\lambda\sigma_3x}
-{\partial\Omega\over\partial
\bar\lambda})\nu^{-1}],
\label{evol-q-gene}\end{equation}
\begin{equation}
{\partial\over\partial t}\ R=[R\ ,\ \Omega]+M.
\label{evol-R-gene}\end{equation}
This general result is not proved here (see \cite{leon-jmp} or
\cite{leon-pla}), but we give in the next section the computation of the time
evolution of $\rho^+(k,t)$ (given here by (\ref{evol-R-gene})) directly from the Lax
pair, when the evolution of $q(x,t)$ is given by (\ref{qt-gene}).

\paragraph*{Solvable evolution.}
The entries $\Omega(k,t)$ (diagonal matrix-valued function)
and $M(k,t)$ (off-diago\-nal matrix-valued distribution) are {\em arbitrary},
which allows to solve evolutions with nonanalytic dispersion
relations ($\partial\Omega/\partial\bar k\ne0$) and {\em arbitrary} boundary
values \cite{leon-pla} (the problem of arbitrary boundary values has been first
solved for the SIT equations in \cite{gabzak}).
As we are interested in solving evolutions as (\ref{qinteq})
where the integral runs on the real $k$-axis, we work within the reduction
(\ref{reduc}) and chose (remember the definition (\ref{star}):
\begin{equation}
M(k,t)=\left(\matrix{0 & m(k,t)\delta^+ \cr -m^*(k,t)\delta^- & 0}\right),
\label{M}\end{equation}
and the dispersion relation
\begin{equation}
\Omega(k,t)=\omega(k,t)\sigma_3,\quad
\omega(k,t)=-{1\over\pi}\int_{-\infty}^{+\infty}{d\lambda\over\lambda-k}\
\varphi(\lambda,t),\quad k\not\in{\bf R},
\label{Omega}\end{equation}
analytic except on the real axis where it stands the discontinuity
\begin{equation}
{\partial\omega(k,t)\over\partial\bar k}=
{i\over2}\left(\omega^+(k_R,t)-\omega^-(k_R,t)\right)\delta(k_I)=
\varphi(k_R,t)\delta(k_I).
\label{d-bar-omega}\end{equation}
Then the quantity $\nu(\partial\Omega/\partial
\bar\lambda)\nu^{-1}$ is ill-defined because $\nu(k,x,t)$ itself is
discontinuous on the real axis. To compute it we need to use the identity 
\cite{marco}
\begin{equation}
\nu\ {\partial\Omega\over\partial\bar k}\ \nu^{-1}=
{\partial\over\partial\bar k}
(\nu\Omega\nu^{-1})-\nu[Re^{2ik\sigma_3x},\Omega]\nu^{-1}.
\label{magic}\end{equation}

Within the reduction and with the above choice of $M$ and $\Omega$, the
evolution (\ref{evol-q-gene}), becomes  by means of (\ref{magic})
\begin{equation}
q_t={2i\over\pi}{\int_{-\infty}^{+\infty}}dk\left[
2\varphi\nu_{11}^+\nu_{12}^-+
m(\nu_{11}^+)^2 e^{-2ikx}+m^*(\nu_{12}^-)^2 e^{2ikx}\right].
\label{qt-gene}\end{equation}
To get the above result, it is necessary to use the Riemann-Hilbert
relations (\ref{R-H}) to express $\nu_{11}^-$ and $\nu_{12}^+$ in terms of
$\nu_{11}^+$ and $\nu_{12}^-$.
This equation, coupled to the system (\ref{lax1}), is the general solvable
evolution with a nonanalytic dispersion law vanishing at large $k$. It is used
now to solve the boundary value problem (\ref{maxsimp})(\ref{qinteq}).

\setcounter{equation}{0}
\section{Solution of SRS}

The above theory allows to solve the nonlinear evolution  problem
(\ref{maxsimp})(\ref{qinteq}) with the  initial data
of $q(x,0)$ (which actually will be taken to vanish) and boundary value
(\ref{in-A}).
This is done in the following way

\paragraph*{Basic relations.}
As the 3 functions
\begin{equation}
\left(\matrix{a_L\cr a_S\exp[2ikx-i\Delta\omega t]}\right),\quad
\left(\matrix{\nu_{11}^+\cr \nu_{21}^+}\right),\quad
\left(\matrix{\nu_{12}^-\cr \nu_{22}^-}\right)\exp[2ikx],
\label{vectors}\end{equation}
solve the same first-order differential system, they are uniquely related
by their values in $x=0$ and comparing (\ref{in-A}) with (\ref{bound-0})
we obtain readily
\begin{equation}
\left(\matrix{a_L\cr a_S\exp[2ikx-i\Delta\omega t]}\right)=J_L
\left(\matrix{\nu_{11}^+\cr \nu_{21}^+}\right)+
J_S\left(\matrix{\nu_{12}^-\cr \nu_{22}^-}\right)\exp[2ikx-i\Delta\omega t].
\label{exp-vect}\end{equation}
This formula actually gives the solution to the physical problem (compute the
output values from the input values) as soon as the function $\rho^+(k,t)$ is
calculated ($\tau$ is expressed from $\rho$ in (\ref{tau-rho})).

\paragraph*{Computation of $\rho(k,t)$.}
This computation is performed by first determining the functions $\varphi(k,t)$
and $m(k,t)$. This is done by equating the evolution (\ref{qinteq}) with 
(\ref{qt-gene})
by means of the relation (\ref{exp-vect}) between $(a_L, a_S)$ and $\nu$.
Using the reduction relations (\ref{red-nu}), 
we finally get that the evolution  (\ref{qinteq}),
where $(a_L, a_S)$ is expressed in terms of $\nu$ through  (\ref{exp-vect}),
reads exactly as  (\ref{qt-gene}) for
\begin{equation}
\varphi=-i{\pi\over4}g_0\left(|J_L|^2-|J_S|^2\right),\quad
m=i{\pi\over2}g_0\ J_LJ_S^*e^{i\Delta\omega t}.
\label{evol-func}\end{equation}

The corresponding evolution (\ref{evol-R-gene}), which reads
\begin{equation}
\rho^+_t=-2\omega^+\ \rho^+-2im,\label{evol-rho-base}\end{equation}
then gives by means of (\ref{Omega})
\begin{equation}
\rho^+_t(k,t)=\pi g_0 J_L(k,t)J_S^*(k,t)e^{i\Delta\omega t}-{i\over2}
g_0\rho^+(k,t)  \int{d\lambda\over\lambda-(k+i0)}\ (|J_L|^2-|J_S|^2) ,
\label{evol-rho}\end{equation}
\begin{equation}
C_{n,t}^+=C_n^+ \left(-{i\over2}g_0\int{d\lambda\over\lambda-k_n}\
(|J_L|^2-|J_S|^2)  \right).
\label{evol-cn}\end{equation}
For everything to work, we have seen in sec. 3 that the analytical requirement
(\ref{rho-analytic}) is essential. It is clear on the above time evolution that
this is ensured for all $t$ if we require that
$J_L(k,t)$ be analytic in the upper half plane and $J_S(k,t)$ in the lower one
(note from (\ref{star}) that $J_S^*(k,t)$ is a function of $k$ analytic in the 
upper half plane). In short
\begin{equation}\label{input-analytic}
{\rm Im}(k)>0\ \Rightarrow\ {\partial\over\partial\bar k}\ J_L(k,t)=0,\quad
{\rm Im}(k)<0\ \Rightarrow\ {\partial\over\partial\bar k}\ J_S(k,t)=0.
\end{equation}
The other constraint is that, for all $t$, the function $\rho^+(k,t)$ vanish at
large $k$ in ${\rm Im}(k)>0$. This is guaranteed here because $J_L(k,t)$,
$J_S^*(k,t)$ and $\omega^+(k,t)$ do vanish at large $k$ in ${\rm Im}(k)>0$.
Note that it would not be true for an analytic dispersion relation
(like  the polynomial $k^2$ for NLS).

A medium initially at rest corresponds to $q(x,0)=0$, and hence from
(\ref{rho+-})  and (\ref{apres1}) to
\begin{equation}
\rho^+(k,0)=0,\quad C_n^+(0)=0.
\label{init-rho}\end{equation}
Consequently the evolution (\ref{evol-cn}) ensures that no bound states are
created and the equation (\ref{evol-rho}) can be solved explicitely. Then the
solution of the related integral equation (\ref{cauchy-1+}) 
(\ref{cauchy-2-}) gives $\nu^\pm(k,x,t)$
which in turn allows to compute the light field envelopes $a_L(k,x,t)$ and 
$a_S(k,x,t)$. 
In particular their values in $x=1$ furnishes the output in the
case of the finite interval of physical length $L$. For instance, with the
solution $\rho^+(k,t)$ of (\ref{evol-rho}), the pump output is obtained by
solving (with $\rho^-=-(\rho^+)^*$):
\begin{equation}
\nu_{11}^+(k,x,t)=1-{1\over2i\pi}\int
{d\lambda\over\lambda-(k+i0)}
\  \rho^-(\lambda,t)\nu_{12}^-(\lambda,x,t)e^{2i\lambda x},
\label{nu-11}\end{equation}
\begin{equation}
\nu_{12}^-(k,x,t)={1\over2i\pi}\int
{d\lambda\over\lambda-(k-i0)}
\  \rho^+(\lambda,t)\nu_{11}^+(\lambda,x,t)e^{-2i\lambda x},
\label{nu-12}\end{equation}
and then it is given by
\begin{equation}
a_L(k,1,t) = J_L(k,t)\nu_{11}^+(k,1,t)
+J_S(k,t)\nu_{12}^-(k,1,t)\exp[2ik-i\Delta\omega t].\end{equation}

We note finally that, in the case of the semi-infinite line ($L\to\infty$), from
the boundary values (\ref{bound-a}), we have the following pump output
\begin{equation}
a_L(k,\infty,t) ={1\over\tau^+}\left(J_L-\rho^+ J_Se^{-i\Delta\omega t}\right),
\label{out-AL}\end{equation}
which has been used in \cite{leon-pra} to interpret the experiments of 
\cite{druwen}.

\setcounter{equation}{0}
\section{Evolution of the spectral transform from the Lax pair}

The auxiliary Lax operator given in (\ref{lax2}) can be simplified by using
(\ref{magic}) and the property that $\nu\Omega\nu^{-1}$ vanishes as ${\cal
O}(1/k)$. Then (\ref{lax2}) reduces to
\begin{equation}
\nu_t=X\ \nu,\quad X=
{1\over2i\pi}\int\!\!\!\int\frac{d\lambda\wedge d\bar\lambda}{\lambda-k}\ 
\nu e^{-i\lambda\sigma_3 x}\left(M+[R,\Omega]\right)
e^{i\lambda\sigma_3 x}\nu^{-1}.
\end{equation}
For the particular choices (\ref{M}) and (\ref{Omega}), the above equation reads
\begin{equation}\label{lax3}
\nu_t=X\ \nu,\quad X(k,x,t)=
-{1\over\pi}\int {d\lambda\over\lambda-k}\ \chi(\lambda,x,t),
\end{equation}
$$
\chi(k,x,t)=(m-i\rho^+\omega^+)
\left(\matrix{-\nu^+_{11}\nu^+_{21} & \nu^+_{11}\nu^+_{11}\cr
-\nu^+_{21}\nu^+_{21} & \nu^+_{11}\nu^+_{21}}\right)\exp[-2ikx]+$$ 
\begin{equation}
+(m^*-i(\rho^+)^*\omega^-)
\left(\matrix{-\nu^-_{12}\nu^-_{22} & \nu^-_{12}\nu^-_{12}\cr
-\nu^-_{22}\nu^-_{22} & \nu^-_{12}\nu^-_{22}}\right)\exp[2ikx].
\label{chi}\end{equation}

Our purpose here is to recover now the time evolution (\ref{evol-rho-base}) of
the reflection coefficient $\rho(k,t)$ by means of the usual method which
consists in evaluating (\ref{lax3}) at one boundary, say in $x=0$. The boundary
values of both $X^+$ and $X^-$ in $x=0$ are easily calculated from those of
$\nu^+$ and $\nu^-$ in (\ref{bound-x=0}). Then the equation (\ref{lax3}) gives
for $\nu^+$ and $\nu^-$ successively
\begin{equation}\label{rho-t-old}
\rho^+_t=
{1\over\pi}\int {d\lambda\over\lambda-(k+i0)}
(i\omega^+\rho^+-m),
\end{equation}
\begin{equation}0=
{1\over\pi}\int {d\lambda\over\lambda-(k+i0)}
(-i\omega^-(\rho^+)^*+\bar m),
\end{equation}
\begin{equation}0=
{1\over\pi}\int{d\lambda\over\lambda-(k-i0)}
(i\omega^+\rho^+-m),
\label{zero-t-old}\end{equation}
\begin{equation}
-(\rho^+)^*_t=
{1\over\pi}\int {d\lambda\over\lambda-(k-i0)}
(-i\omega^-(\rho^+)^*+\bar m).
\end{equation}

First, these equations are compatible if $\omega^-=(\omega^+)^*$, which is
guaranteed through (\ref{Omega}) by $\varphi\in i{\bf R}$ (see
(\ref{evol-func})). Then the above four equations reduce to two and we select
(\ref{rho-t-old}) and (\ref{zero-t-old}) whose solution goes in two steps.\\
{\em First step}: the equation  (\ref{zero-t-old}) is realized if the function
$i\omega^+\rho^+-m$ is analytic in Im$(k)>0$ and vanishes as
$k\to\infty$. This is true by construction for $\omega(k,t)$ and
this is true also for $m(k,t)$ because we have required that
$J_L(k,t)$ be analytic in the upper half plane and $J_S(k,t)$ in the lower one
(rmember that  from (\ref{star}) that $J_S^*(k,t)$ is a function of $k$ analytic
in the  upper half plane). Finally this holds also for the reflection coefficient
$\rho^+(k,t)$  by the requirement (\ref{rho-analytic}).\\
{\em Second step}: by substraction of (\ref{rho-t-old}) and (\ref{zero-t-old}),
and by use of the Sokhotski formula (\ref{sokhotski}) we arrive precisely at the
time  evolution (\ref{evol-rho-base}).

Finally, using the behaviors in $x=\infty$ instead of $x=0$, furnishes the time
evolution of the transmission coefficient $\tau^\pm$.

{\bf  AKNOWLEDGEMENTS}

It is a pleasure to thank M. Boiti, A.V. Mikhailov and F. Pempinelli
for enlighting discussions and constructive comments.

\end{document}